\documentstyle[epsfig,12pt]{article}

\begin{document}
\begin{center}
{\large\bf Tensor analyzing powers in  forward angle exclusive $\pi$-meson 
photoproduction on  deuteron.}\\ 

\vspace{4mm}

{ A.Yu. Loginov,  A.V. Osipov,  A.A. Sidorov, V.N. Stibunov. }\\
{\small \em Nuclear Physics Institute at Tomsk Polytechnic University, 634050 Tomsk, Russia}
\end{center}

\vspace{6mm}

\begin{abstract}
The target tensor analyzing powers of the process   $\gamma d\rightarrow \pi^{-}pp$  had been studied
in the plane wave impulse approximation, if  both ejected protons are detected in coincidence and in
the directions,  symmetrical respect to the incoming photon momentum, being in the proton plane.
The matrix elements  of the studied reaction become essentially  simple, if the magnitudes of  the proton
momenta $ p_1=p_2=p $ are equal. That proton kinematics keeps the ejected pion angle near zero, therefore
the one body pion production operator  does not include  the spin-non-flip term.  Moreover, the transition
to the triplet, spatially antisymmetric final $pp-$ state  is strongly suppressed  from  the symmetrical
deuteron $ S -$ state as well as matrix elements have  a large dependence from the orientation  of 
the deuteron spin. The values of the $T_{22}(\vec p)$ component,  calculated with three  realistic deuteron wave 
functions  in the proton momentum region $p \ge 350 MeV/c$ and proton polar angle $\theta \ge 40^{\circ}$,
differ one from another noticeably.        
\end{abstract}
\begin{quotation}
   
Meson photoproduction from nuclei on a level with elastic and inelastic electron
scattering for a long time is an important reaction used for the investigation of the nuclear
 structure and dynamics. To examine  the pion photoproduction on the neutrons,that is
 great particular interest, as a rule, the investigation of the reaction $\gamma d\rightarrow \pi
 NN $  is being used.
However the extraction of the information even on the simplest  nucleus structure
as well as on pion production on a neutron is very complicated by uncertainties aroused
from initial and final state interaction, Fermi motion, off-mass-shell effects. Moreover
for the charge pion production exclusion principle effect on the final $nn$ or $pp$ system
should be taken into consideration. The experimental and theoretical studies of the pion 
photoproduction reactions involving polarized deuteron target have received considerable 
attention at present. It is apparent that the polarization observables may provide a means of obtaining 
additional nuclear dynamics information.
 
We have studied  experimentally the differential cross section and tensor target asymmetries of the process
\begin{equation}
\gamma d\rightarrow \pi^{-}pp                                 
\end{equation}
by recording two protons in coincidence  in the reaction 
\begin{equation}
ed\rightarrow  pp \pi^{-}e^{\prime}                            
\end{equation}
at the photon point, using  the internal polarized deuterium target of the VEPP-3 storage ring \cite{Log}.
The experimental data have been obtained at large ejected proton momenta 
$ 300 {MeV}/{c} < p < 700{MeV}/{c} $.
The detector angular acceptance was $ 63^{\circ} < \theta <83^{\circ}, \Delta\varphi=40^{\circ} $.
As the experimental yield  proved to be not very prosperous, we have to divide the whole data into 
bins, corresponding to a range division of one in a six independent parameters only. So the 
presented differential cross section and asymmetry components had been effectively integrated 
over five remaining parameters in their allowed boundaries. The resulting differential cross section 
and asymmetry components differed significantly from the results obtained in a plane wave impulse 
approximation (PWIA).
 
The recent example of the  theoretical investigation of the single spin 
asymmetries for the inclusive reactions $ d(\gamma,\pi) NN $  is \cite{Dar}. One can arrive at
a conclusion that measurements of $T_{2M}$ would not allow
us discriminate between physically reasonable deuteron wave function (DWF) with different
percentage D - state. Actually, the tensor analyzing powers, obtained with DWF of the Bonn
potential\cite{Mac} instead of the one of the Paris potential\cite{Lac} had a minor differences.

The matrix element of the process (1) in PWIA  is given by \cite{Dar}:
\begin{eqnarray}
M_{smm_{\gamma}m_{d}}(\vec k,\vec q,\vec p_{1},\vec p_{2}) = \sqrt2\sum_{m^{\prime}}< sm| 
[< \vec p_{1}|t(\vec \varepsilon_{m_{\gamma}},\vec k, \vec q)|- \vec p_{2}> \psi_{m^{\prime}m_{d}}
(\vec p_{2}) \nonumber\\+ (-1)^{s} < \vec p_{2}|t(\vec \varepsilon_{m_{\gamma}},\vec k, \vec q)|- \vec p_{1}>
\psi_{m^{\prime}m_{d}}(\vec p_{1})]|1m^{\prime}> ,                               
\end{eqnarray}
where $s$, $m - $ total spin and projection of the two outgoing protons,
\begin{equation}
\psi_{m^{\prime}m_{d}}(\vec p) = (2\pi)^{3/2}\sqrt{ 2E_{d}}\sum_{L=0,2}\sum_{m_{L}}
i^{L}<Lm_{L}1m^{\prime}|1m_{d}>u_{L}(\vec p)Y_{Lm_{L}}(\hat{\vec p})\chi_{1m_{d}}                                 
\end{equation}
is the deuteron wave function  in the momentum space, $ \hat{\vec p}=\vec p/|\vec p|$ .
$Z -$ axis of our right - handed coordinate system is directed along the momentum $ \vec k$ of the
incoming photon, and $ y - $ axis along $\vec p_{2} \times \vec p_{1}$.
 
Matrix element (2) possesses the symmetry under interchange of the proton momenta, so the 
elementary operator $ t(\vec \varepsilon_{m_{\gamma}},\vec k, \vec q)$ acts on the nucleon with
number 1 only. This operator has a general form \cite{Ch}:
\begin{equation}
 t(\vec\varepsilon_{m_{\gamma}},\vec k, \vec q) = i\vec \sigma \vec \varepsilon_{m_{\gamma}} F_1
 +(\vec \sigma \hat{\vec q}) (\vec \sigma [\hat{\vec k}\times\vec \varepsilon_{m_{\gamma}}]) F_2  
 +i (\vec \sigma \hat{\vec k}) (\vec \varepsilon_{m_{\gamma}}\hat{\vec q}) F_3     
 +i (\vec \sigma \hat{\vec q}) (\vec \varepsilon_{m_{\gamma}}\hat{\vec q}) F_4 .                            
\end{equation}
Here $\hat{\vec k} = \vec k/|\vec k|  ,  \hat{\vec q} =\vec q/|\vec q|$ and $ \vec\varepsilon_{m_{\gamma}}$
is photon polarization vector of helicity $m_{\gamma}$.

As can be seen from equation of the reaction amplitude (3), 
every measurable variables of the reaction (1) must be a complex function of the two general
factors. One of them is determined by meson photoproduction process on a nucleon, while the 
other by nuclear structure properties. 

The simple spectator model  is attractive because there is a hope to connect the experimental data
analysis of the pion  photoproduction on deuteron target with  DWF in the momentum
representation by the most direct way.  However slow nucleon for example $p_{1}$ can be identified 
as the spectator, and $\psi_{m^{\prime}m_{d}}(\vec p_{2})$ can be completely neglected, if the other
nucleon is much faster $p_{2}\gg p_{1}$. When $p_{1}$ and $p_{2}$ is comparable, correct analysis is
essentially more complicate for extracting the free - nucleon data \cite{Dea}.  

We have improved the agreement   between the calculation  results and experimental data \cite{Log} 
taking into consideration effects of the final state interaction, but  we were unable to study the nucleon
momentum distribution in the deuteron from a relatively  direct comparison of the experimental data
about the components of  tensor target asymmetry  with results of the calculations. It is not only because
the elementary pion production  operator includes the many contributions, but also because the necessity
to symmetrize the final  $pp$ - state to take into account the Pauli blocking.  In \cite{Dea} it had been estimated 
that  effect of  symmetrization is to enhance the spatially symmetric final states, up to a factor 4, and
to suppress correspondingly the antisymmetric ones. 

As a result of symmetrization, the analyzing power components $T_{2M}$ even in PWIA are  expressed 
in the numerous terms, including  varied products of the bilinear combinations of the S-  and  D - waves 
of the DWF and the different parts of  the elementary amplitude.  Effect of the rescattering on polarized 
observables of the reaction (1) in the Delta - resonance region was studied in \cite{Log1}.  
The theoretical  analysis  of the obtained expressions is complicating infinitely. 
 
The first major objective of this contribute is to determine which measurements
are valid for extraction essential information on the free - nucleon data or on nucleon momentum
distribution in the deuteron.  We have paid attention to the possibility to derive the advantage
from the symmetrization of the proton-proton state when both proton momenta are under control.
The single pion photoproduction (1) is described by six independent variables, if the incoming photon has
energy spectrum. In accordance with the experimental kinematics \cite{Log} the proton momenta $p_{i}$,
their angles $ \theta_{i}$ and $\varphi_{i} $ (i=1,2) has been chosen as independent variables. From 
the momentum and energy conservation one can obtain the photon energy quantity.
First of all, if the both ejected protons are observed in the directions, symmetrical respect to the incoming
photon momentum, being in the proton plane, i.e. $ \theta_{2} \approx \theta_{1}, \varphi_{2} \approx 
\varphi_{1} + \pi $ and the magnitudes of the proton momenta are near to each other, the matrix
elements of the reaction (1) become simpler essentially.
 
The point is that above proton kinematics keeps the ejected pion angle near zero, therefore the one body
pion production operator, that is the important ingredient of the studying of the pion production on the
deuteron, does not include  the spin-non-flip term. 
The elementary pion production operator $ t(\vec\varepsilon_{m_{\gamma}},\vec k, \vec q)$ will preserve
the first term $ i\vec \sigma \vec \varepsilon_{m_{\gamma}} F_1$ only. As a consequence, the transition to
the triplet, spatially antisymmetric  final $pp-$ state  is strongly  suppressed  for  the symmetrical deuteron
$ S -$ state  as well as matrix elements   have a large dependence  from the orientation  of  the deuteron spin. 
Polarization studies are seen   to be of the considerable  practical  importance in these investigations. 
 
As we have analyzed, the square of the matrix element of  the reaction (1)  after summing over 
the final spin states and averaging over the photon polarizations for $m_{d}=0$
is proportional to the $D-$wave of DWF $w(\vec p)^{2}$ only. However the experimental measurements
of the cross sections of the process (1), using tensor polarized deuterium target with $ m_d = 0$, seemed
quite unlikely,  because  such high value of the target  polarization for the present is unrealizable.
 
In this paper we shall concentrate on the investigation of the tensor analyzing power components
in negative  pion photoproduction from deuteron:
\begin{eqnarray}   
T_{2M}=\frac{ Sp M \tau_{2M} M^{+}}{  Sp M M^{+} },
\end{eqnarray}
 where $\tau_{2M}$ is the spherical spin - tensor in the deuteron matrix density decomposition.
 
 The elementary pion photoproduction operator does not play any role in the target asymmetry
calculations for our symmetrical kinematics because it produces the equal factors in both a numerator and
a denominator of (6), so  the $ T_{2M}$ is expressed  in terms of the  $ S - $ and $ D - $ wave of the DWF.
 
\begin{eqnarray}   
T_{20}=\frac{32\sqrt{2}u^2(p)-16(3cos(2\theta)+1)u(p)w(p)+\sqrt{2}(-12cos(2\theta)+9cos(4\theta)+19)w^{2}(p)}
 {4(16u^2(p)-4\sqrt{2}(3cos(2\theta)+1)u(p)w(p)+(-6cos(2\theta)-9cos(4\theta)+23)w^{2}(p))},
\end{eqnarray}
\begin{eqnarray}
T_{21}=0,\nonumber\\   
\end{eqnarray}
\begin{eqnarray}   
T_{22}=-\frac{3\sqrt{3}sin^2(\theta)w(p)(4\sqrt{2}u(p)+3cos(2\theta)+5)w(p)}
 {16u^2(p)-4\sqrt{2}(3cos(2\theta)+1)u(p)w(p)-(6cos(2\theta)+9cos(4\theta)-23)w^{2}(p))}.
\end{eqnarray}
We have estimated tensor analyzing power components $T_{20} , T_{22}, $using DWF 
of Bonn \cite{Mac}  and Paris\cite{Lac} potential model and one, obtained in \cite{Cer}, corresponding
to a percentage D state $ P_{D}=8\%$: Figs. 1.- 4.
The $ T_{22} $, as a function of the proton polar angle, calculated using the DWF of Bonn potential differs
essentially from it, calculated with DWF of Paris potential, starting from $ p = 350 MeV/c$. The
magnitude of $ T_{22} $ in that region is small, and reaches higher values with increase of the proton
momentum: Fig.2., middle and bottom panels. This choice effect for $ T_{20} $ is weaker, but still
noticeable particularly for proton polar angles near $ 30^{\circ} $ at proton momenta $ p \geq 400 MeV/c$ .   

\begin{figure}
\includegraphics[width=0.47\textwidth]{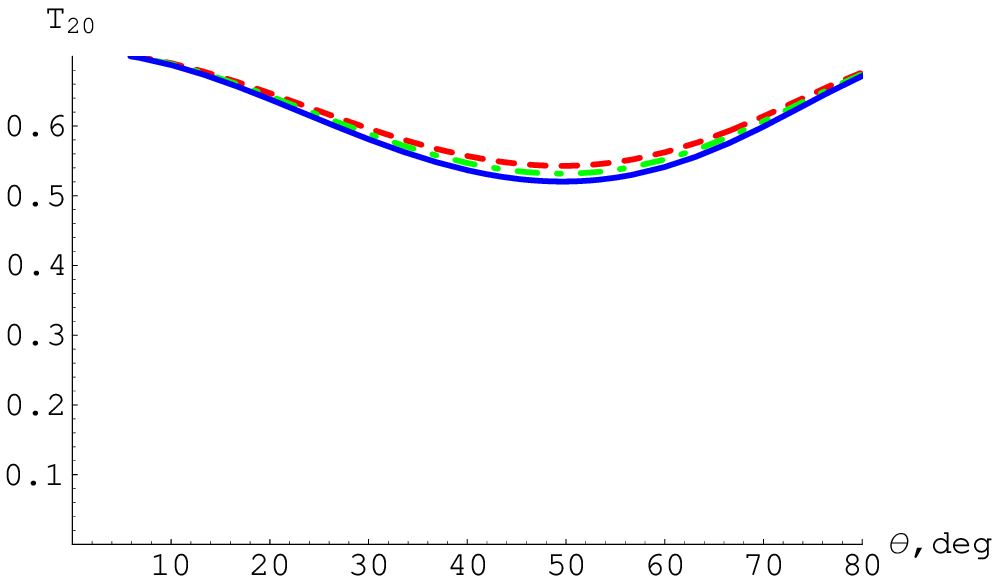}
\hfill
\includegraphics[width=0.47\textwidth]{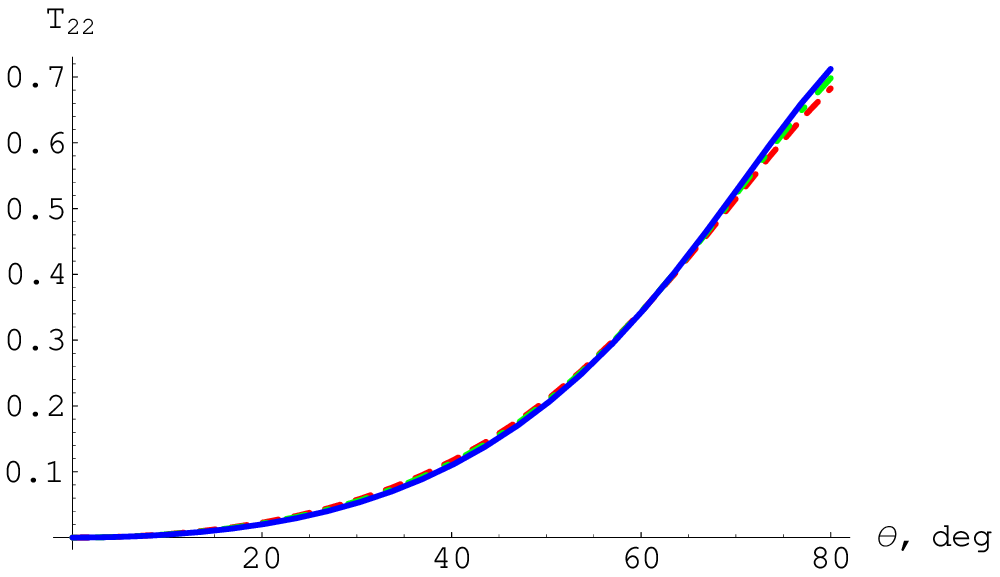}
\includegraphics[width=0.47\textwidth]{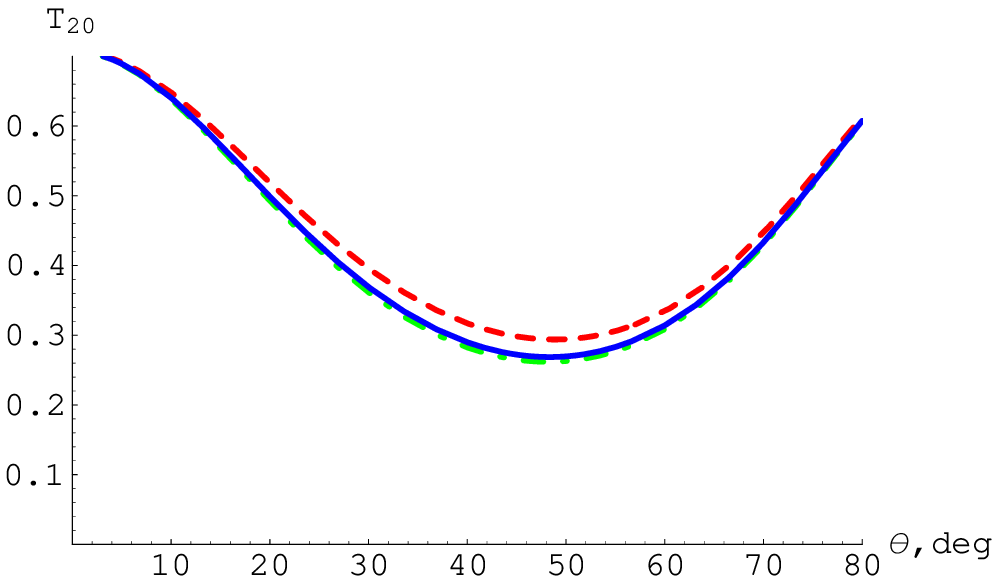}
\hfill
\includegraphics[width=0.47\textwidth]{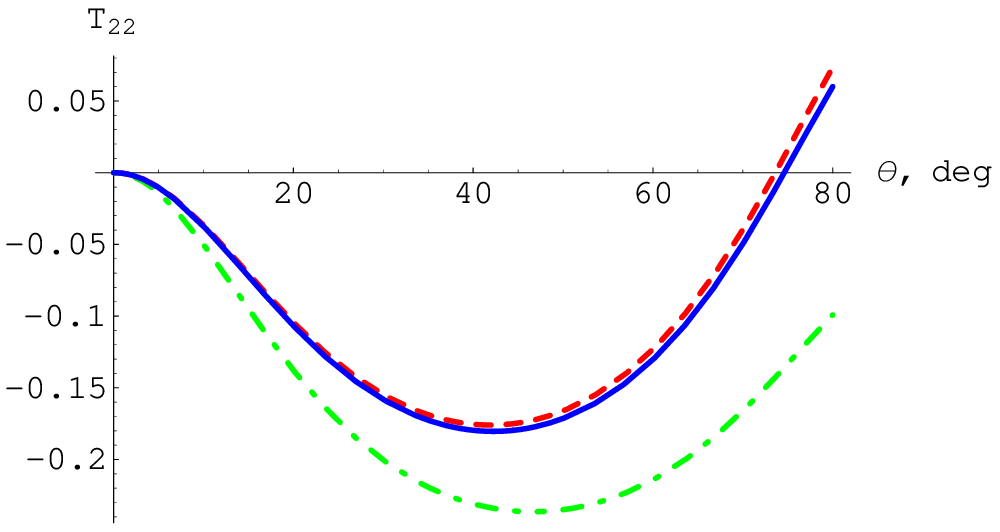}
\includegraphics[width=0.47\textwidth]{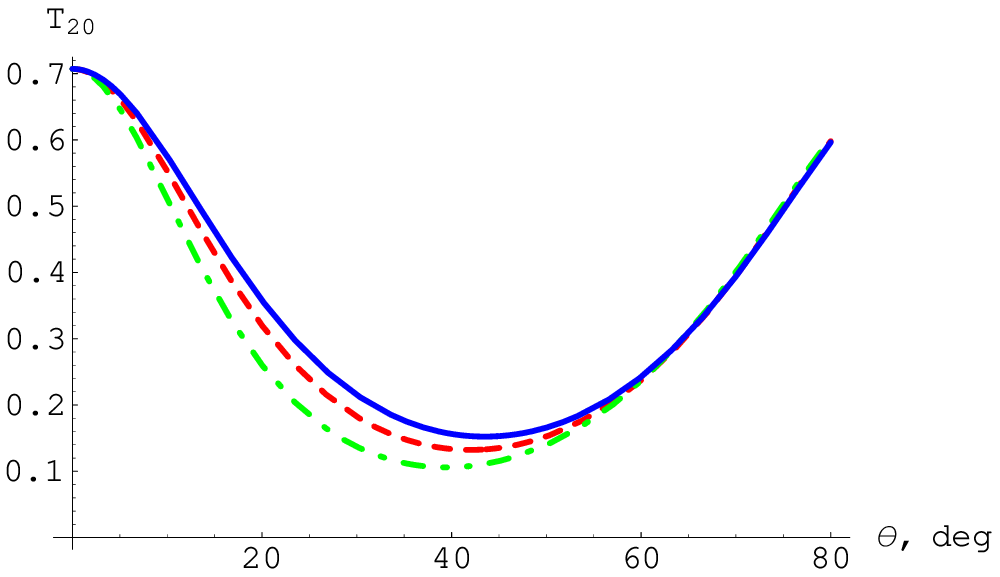}
\hfill
\includegraphics[width=0.47\textwidth]{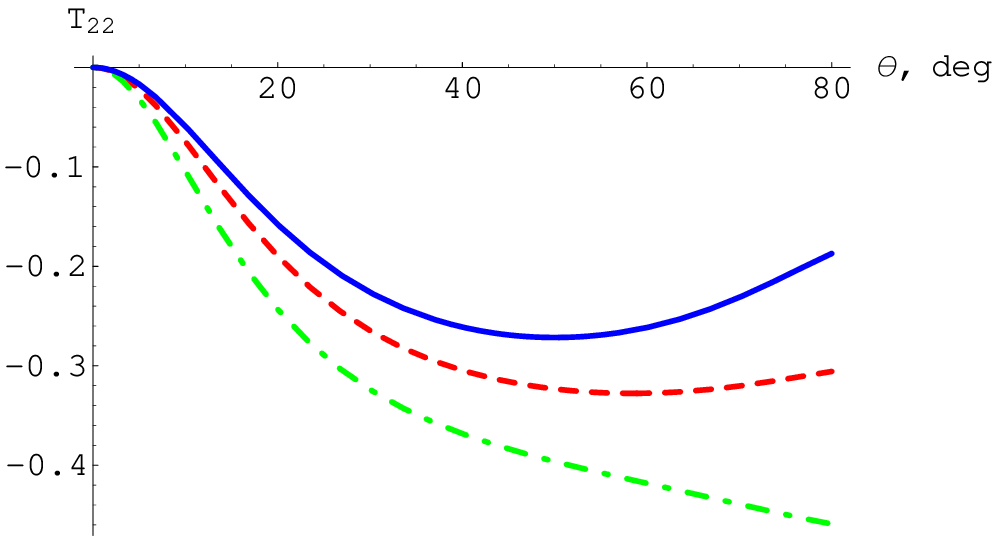}
\\
\parbox[t]{0.47\textwidth}{\caption{T20 component of the tensor analyzing power as a function 
of the proton angle $\theta$ at the different proton momenta: top -- $p=200MeV/c$,  
middle -- $p= 300MeV/c$,  bottom  -- $p= 400MeV/c$. Dashed (dash - dotted) curves --  calculation 
using the DWF of Bonn\protect\cite{Mac} (Paris \protect\cite{Lac}) potential model. Solid line -- calculation using
the DWF parametrization of \protect\cite{Cer} with $P_{D} = 8\%$.}\label{f:1}}
\hfill
\parbox[t]{0.47\textwidth}{\caption{T22 component of the tensor analyzing power as a function 
of the proton angle  $\theta$ at the different  proton momenta: top -- $p=200MeV/c$,  
middle -- $p= 350MeV/c$,  bottom -- $p= 400MeV/c$. Notation of the curves  as in Fig. 1.}
\label{f:2}}
\end{figure}
    
If ejected proton angles $\theta$ obey the condition $1+3cos(2\theta)=0$, that is equivalently to
$3cos^2(\theta)-1 =  0 $, i.e. $Y_{20}(\theta)=0$ the expressions, obtained for $T_{20}$ and $T_{22}$ in the 
impulse approximation are as follows:               
$$T_{20}=\frac{2u^2(p)+w^{2}(p)}{2\sqrt{2}(u^2(p)+2w^2(p))} ,  T_{21}=0,    T_{22}=-\frac{\sqrt{3}w(p)
(\sqrt{2}u(p)+w(p))}{2(u^2(p)+2w^2(p))}.$$

\begin{figure}
\includegraphics[width=0.47\textwidth]{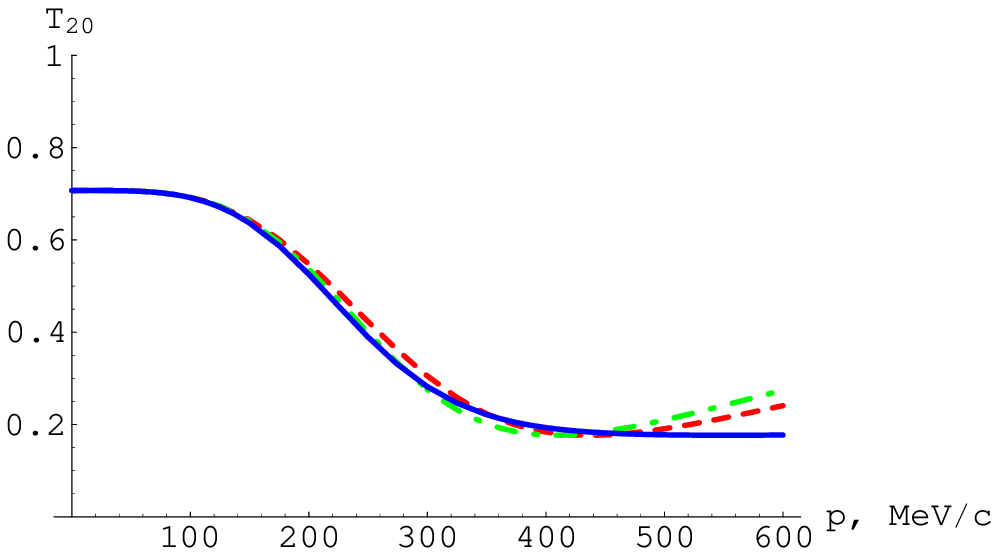}
\hfill
\includegraphics[width=0.47\textwidth]{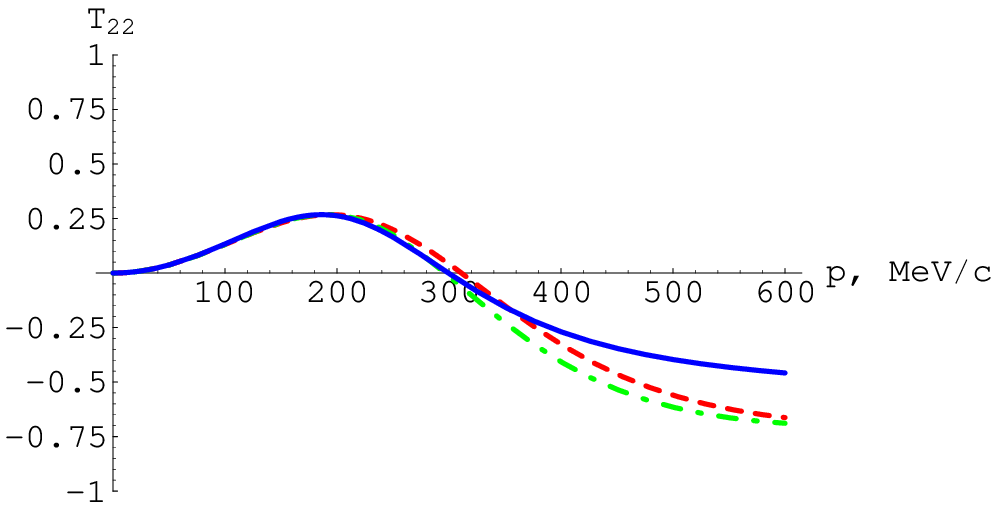}
\\
\parbox[t]{0.47\textwidth}{\caption{T20 component as a function of the proton momentum $p$. 
Notation of the curves as in Fig. 1.}\label{f:3}}
\hfill
\parbox[t]{0.47\textwidth}{\caption{T22 component as a function of the proton momentum $p$.
Notation of the curves as in Fig. 1.}\label{f:4}}
\end{figure}

Figs. 3,4 show the momentum dependence of the $ T_{20}$ and $ T_{22}$ components of the tensor analyzing
power. Whereas the $ T_{20}$ curves, calculated with three DWF, do not differ  up to $ 500 MeV/c$,
the $ T_{22}$, calculated with Bonn and Paris DWF, differing one from another only a little, exceed in the absolute
value result of the calculation with DWF \cite{Cer} essentially.

As an example we have received in one value of the analyzing power components $ T_{20}, T_{22} $ with
obviously poor statistical accuracy, introducing the restrictions on our early experimental data \cite{Log}
$ | \vec p_{1}-\vec p_{2} | \leq 50 MeV/c,  |\theta_{1}-\theta_{2}|\leq 3.5^{\circ} $    
$ T_{20}(\langle p\rangle = 400 MeV/c,\langle \theta \rangle = 71.5^{\circ}) = 0.68 \pm 0.65,
 T_{22}(\langle p \rangle = 400 MeV/c,\langle \theta \rangle = 71.5^{\circ}) = - 0.16 \pm 0.53 $ .
These values do not contradict our predictions.

 We shall examine the possibilities of the new experimental data from $\gamma \vec{d}\rightarrow \pi^{-}pp - $
reaction to study in this frame $w(p)$  and the ratio $u(p)/w(p)$  for the different realistic DWF. 
The  measurements of the $T_{22}$ angular dependence in the region $p \geq 400MeV/c$ with high precision, 
in principle, can allow us to select the most suitable DWF of them,  having different $P_{D}$ values. 

The main approximation in this work is the neglect of the final state interaction. So, the differences of the 
measured $T_{2M}$ from they, predicted by formulae (7), (8) must be explane by the necessity to consider
more complicated mechanisms of the reaction (1), and this difference will determine the contributions of the 
pion - nucleon, nucleon - nucleon rescattering, meson exchange currents  to the amplitude.

This work was supported by the Russian Foundation for Basic Researches, grants 01-02-17276 
and 04-02-16434, by RF Ministry of Education, grant E02 -3.3-216 and by SNT Program "Nuclear Physics".
\end{quotation}

\end{document}